\documentstyle[aps,epsfig,preprint]{revtex}

\begin{document}

\tighten 

\title{Quasi-Deuteron Configurations in odd-odd N=Z nuclei} 
\author{A.~F.~Lisetskiy$\,^1$, R.~V.~Jolos$\,^{1,2}$, N. Pietralla$\,^1$, 
 P. von Brentano$\,^1$\\}

\address{$^1\,$ Institut f\"ur Kernphysik, Universit\"at zu K\"oln, 
                50937 K\"oln, Germany \\
         $^2\,$ Bogoliubov Theoretical Laboratory, 
        Joint Institute for Nuclear Research, \\
        141980 Dubna, Russia} 

\date{\today}

\maketitle

\begin{abstract}
  The isovector M1 transitions between low-lying T=1 and T=0 states in 
  odd-odd N=Z nuclei are analyzed. Simple analytical expressions for 
  M1 transition strengths are derived within a single-j-shell approximation 
  for both $j=l+1/2$ and $j=l-1/2$ cases. The large B(M1) values for 
  the $j=l+1/2$ case are attributed to {\em quasi-deuteron configurations}.
  The B(M1) values for the $j=l-1/2$ case are found to be small due to 
  partly cancellation of spin and orbital parts of the M1 matrix element.  

\end{abstract}


\section{Introduction}

The structure of selfconjugate nuclei with equal numbers of protons (Z) and 
neutrons (N=Z) is currently attracking a lot of attention. 
The structure 
of N=Z nuclei provides a sensitive test for the isospin symmetry 
\cite{Heiz32} of nuclear forces. 
It is well known that the structure of even-even nuclei with protons and
neutrons occupying different shells is determined by Cooper type pairs
with isospin T=1 and angular momentum J=0 formed by nucleons of the same kind. 
In nuclei close to the N=Z line, where valence protons and valence 
neutrons occupy the same shells in addition to the standard 
pair correlations mentioned above proton-neutron pair 
correlations with $T=1$ and with $T=0$ can become important.
It means that in addition to the proton-proton and neutron-neutron 
$0^+$ pairs proton-neutron pairs with different angular momenta 
can play an important role. 
Below we will analyse experimental information, which demonstrates the 
important role of the neutron-proton pairs with $J^{\pi}_T=1^+_0$  
and $J^{\pi}_T=0^+_1$ in the structure of N=Z nuclei. 
In $N=Z$ nuclei both kind of states with total isospin quantum numbers 
$T=0$ and $T=1$ exist. 
In odd-odd N=Z nuclei the lowest T=0 states and T=1 states are 
low-lying (below 4 MeV).
This unique phenomenon is in contrast to even-even N=Z 
nuclei, where the T=0 $0^+_1$ ground state is lowered and is separated from 
excited T=1 states by a large energy gap. A lot of work, experimental  
\cite{Sve98,Sko98,Ang98,Rudolph,Exp1,Ter98,Vin98,Lenzi98,Fries98,Schneider99} and theoretical
\cite{Faes80,Faes99,Lang96,Isa97,Sat97,Buc97,Ots98,Cast91,Ron98,Zamick,Fuji97,Poves}, has been carried out recently for the investigation and understanding of 
the N=Z nuclear structure. One interesting phenomenon is the occurrence of 
very large magnetic dipole (M1) matrix elements between nuclear states along
the N=Z line. 
The M1 moments of odd-odd N=Z nuclei have recently been revisited
\cite{Ron98} within a simple shell model approach. Another recent work 
\cite{Zamick} discusses the interference term between spin and orbital 
contributions to M1 transitions in even-even s-d shell nuclei. 

In the present article we would like to focus on isovector M1 transitions 
strength between low-lying states in odd-odd N=Z nuclei which are accessible
to $\gamma$-spectroscopy.
Some of the odd-odd N=Z nuclei exhibit very strong isovector M1 transitions 
between the yrast states with quantum numbers $J^\pi_T=0^+_1$ and $1^+_0$ 
(see Table \ref{m1values}). In a few nuclei, $^{10}$B,$^{22}$Na and 
$^{26}$Al, the strong M1 transitions are fragmented among two or three states.
Other odd-odd N=Z nuclei $^{14}$N, $^{30}$P, $^{34}$Cl, $^{38}$K have 
considerably weaker $0^+_1 \rightarrow 1^+_0$ transitions, in some cases 
with almost vanishing M1 strengths. The total 
B(M1;$0^+_{T=1} \rightarrow 1^+_{T=0}$) strength between the low-lying states 
in odd-odd N=Z nuclei depends sensitively on the mass number A and do not 
show a smooth behavior due to the underlying nuclear shell structure. 

Nowadays exact shell model calculations can be performed for nuclei 
with mass numbers smaller than about $Aß\approx 60$ with 
conventional techniques, e.g. \cite{Poves}. 
The shell model problem can be, at least partly, solved 
approximately, but with controlable accuracy, for heavier nuclei 
with newly developed statistical Monte Carlo methods 
\cite{Lang96,Ots98}. 
Both numerical techniques are powerful and important methods 
to describe in detail the structure of light and medium mass 
nuclei, including $M1$ properties of $N=Z$ nuclei, 
on a microscopic level. 
Besides these sophisticated numerical approaches sometimes 
simple models can yield analytical results, which can help 
to clarify the underlying physics of a certain phenomenon 
in an approximate but simple and transparent way. 
Important examples are the analytical Schmidt values for 
magnetic moments of odd-mass nuclei. 
The Schmidt values were obtained in a pure, i.e. non-interacting, 
shell model approach considering one nucleon outside the even-even 
core in a single-$j$ shell orbital. 
The Schmidt values serve as important benchmarks for the 
actual values of $M1$ moments found experimentally in 
odd-$A$ nuclei. 

In the next section we will discuss analytical formulae 
for $T=0 \to T=1$ isovector $M1$ transition matrix elements 
in odd-odd $N=Z$ nuclei, which we derive in a simple 
core+two-nucleon single-$j$-shell approximation. 
M1 transition matrix elements are found to be 
large between two-nucleon {\em quasi-deuteron configurations}, which we 
define below. Reduction of M1 strengths in other cases will be understood as 
partly cancellation of spin and orbital parts of M1 matrix elements. 
Analytical expressions will be given, which relate the isovector M1 
transition strengths in N=Z odd-odd nuclei to magnetic moments in neighboring 
odd-mass nuclei. 
In section 3 we will compare the experimental data to the simple analytical 
formulae. Good agreement is obtained. Predictions for isovector M1 transition 
strengths in heavier odd-odd N=Z nuclei are done from an extrapolation of 
our formulae up to $^{82}$Nb.

\section{Analytical Formulae for M1 transition strengths}

Attempting to understand the observed data on 
isovector $0^+_{T=1} \rightarrow 1^+_{T=0}$ M1 transition strengths 
in odd-odd N=Z nuclei we have applied the shell model 
in the core+{\bf two-nucleon 
single-j}-shell approximation (2NSj). 
I.e., we consider an odd-odd N=Z nucleus as an 
inert $J^\pi_T=0^+_0$ even-even N=Z core with two valence nucleons, 
one proton and one neutron, in the same shell model orbital 
with quantum numbers (nlj). 

These two valence nucleons can couple to product states with 
total angular momentum 
$J=0,1,...,2j$ and positive parity. The states with even spin quantum numbers
have the isospin quantum number T=1 and states with odd J have T=0. The free
one-proton--one-neutron system is the deuteron. In the lowest states of the 
deuteron, the bound $J_T^\pi=1^+_{0}$ ground state and the unbound  
$J_T^\pi=0^+_{1}$ resonance, both nucleons occupy 
the $1s_{1/2}$ shell with $j=l+1/2$. 

In generalization of the deuteron case we denote the wave functions in the 
2NSj approximation as {\em quasi-deuteron configurations} (QDC) in the 
$j=l+1/2$ cases. This is in agreement with the conclusion made in 
\cite{Ron98}. It is clarified below that M1 properties of QDC differ 
considerably from the case with $j=l-1/2$ due to the interference of orbital 
and spin parts in the M1 matrix elements. 
In reality the proton-neutron pairs coupled to angular momentum 
$J^{\pi}=0^+$ or $1^+$ can be distributed with some weights over 
the several single particle (nlj) states. Experimental 
indications on this effect will be also briefly discussed below.   

 The M1 transition operator
\begin{equation}
\label{op}
{\bf T}(M1)=\sqrt{3\over 4\pi}\left[g_p^l{\bf L}_p+g_p^s{\bf S}_p+g_n^l{\bf L}_n+g_n^s{\bf S}_n\right]{\mu_N \over \hbar} 
\end{equation}
is the sum of the orbital and spin parts for protons and for neutrons.
Here ${\bf L}_\rho ({\bf S}_\rho$) is the orbital (spin) angular momentum
operator for $\rho \in \{p,n\}$, $g_{\rho}^{l(s)}$ is the orbital (spin) 
g-factor and $\mu_N=e\hbar/2M_pc$ represents the nuclear magneton.  

 A $\Delta$T=1 isovector M1 transition, for instance between the $0^+_{T=1}$ 
 and $1^+_{T=0}$ yrast states in odd-odd N=Z nuclei, is generated by the 
isovector (IV) part of the M1 transition operator 
\begin{equation}
\label{iv}
{\bf T}_{IV}(M1)=\sqrt{3\over 4\pi}\left[{g_p^l-g_n^l \over 2}({\bf L}_p-{\bf L}_n)
+{g_p^s-g_n^s \over 2}({\bf S}_p-{\bf S}_n)\right]{\mu_N \over \hbar}. 
\end{equation} 
This is a consequence of the tensor properties of the M1 transition operator 
in the isospin space. In the simple 2NSj approximation it is possible 
analytically to derive expressions for the reduced matrix elements (m.e.) of 
the M1 transition operator. 
For the M1 transition m.e. between states with total angular momentum quantum 
numbers J=0 and 1 , i.e.,     
$\langle (\pi j \otimes  \nu j); J=0\|{\bf T}(M1)\| (\pi j \otimes \nu j); J=1 \rangle$, 
one obtains  :
\begin{eqnarray}
\label{me1}
\langle0^+\|{\bf T}(M1)\|1^+ \rangle = \sqrt{{3\over 4\pi}{j+1\over j}} 
\left[(g_p^l-g_n^l)l+{g_p^s-g_n^s \over 2}\right]\mu_N,\hspace{0.3cm} \mbox{for}\hspace{0.3cm}  
j=l+{1 \over 2},  
\end{eqnarray}
 i.e. for QDC, and  
\begin{eqnarray}
\label{me2}
\langle 0^+\|{\bf T}(M1)\|1^+ \rangle = \sqrt{{3\over 4\pi}{j\over j+1}} 
\left[(g_p^l-g_n^l)(l+1)-{g_p^s-g_n^s \over 2}\right]\mu_N,
\hspace{0.3cm} \mbox{for}\hspace{0.3cm}  
j=l-{1 \over 2}. 
\end{eqnarray} 
The analysis of the Eq.(\ref{me1}) and (\ref{me2}) results in some 
interesting conclusions. Comparing Eq.(\ref{me1}) and Eq.(\ref{iv}) one can 
see that the orbital proton $\langle{\bf L}_p\rangle$ and neutron 
$\langle{\bf L}_n\rangle$ nondiagonal m.e. have opposite signs and equal 
absolute values. This is valid also for the spin proton 
$\langle{\bf S}_p\rangle$ and neutron $\langle{\bf S}_n\rangle$ m.e.. Since 
the  $\langle{\bf S}_p\rangle$ and $\langle{\bf S}_n\rangle$ have 
opposite signs as well as spin $g_p^s$ and $g_n^s$ factors, the nondiagonal 
spin part of the total m.e. of the isovector M1 transition operator is large 
for both the $j=l + 1/2$ and the $j=l - 1/2$ cases. The orbital part of the 
total m.e. 
increases with increasing $l$ and can be comparable with the spin part for   
both cases. However, in the case of $j=l + 1/2$ i.e., when the orbital 
angular momentum and spin of the single particle are aligned, the spin and 
orbital parts are summed up in phase, which results in large absolute values 
of the total m.e. and consequently in large values of the reduced M1 
transition strength 
$B(M1;0^+_1 \rightarrow 1^+_0)=\langle 0^+\|{\bf T}(M1)\|1^+ \rangle^2$. 
The opposite happens in the $j=l-1/2$ case -- the orbital 
and spin parts partially cancel and the reduced M1 m.e. becomes 
small. The constructive $[$Eq.(\ref{me1})$]$ and destructive 
$[$Eq.(\ref{me2})$]$
interference of the orbital and spin parts plays also an important 
role in Gamow-Teller transitions and M1 $\gamma$-transitions in 
even-even N=Z nuclei, as it was recently shown in \cite{Zamick,Fuji97}.

\begin{figure}[bt]
\centerline{\epsfig{file=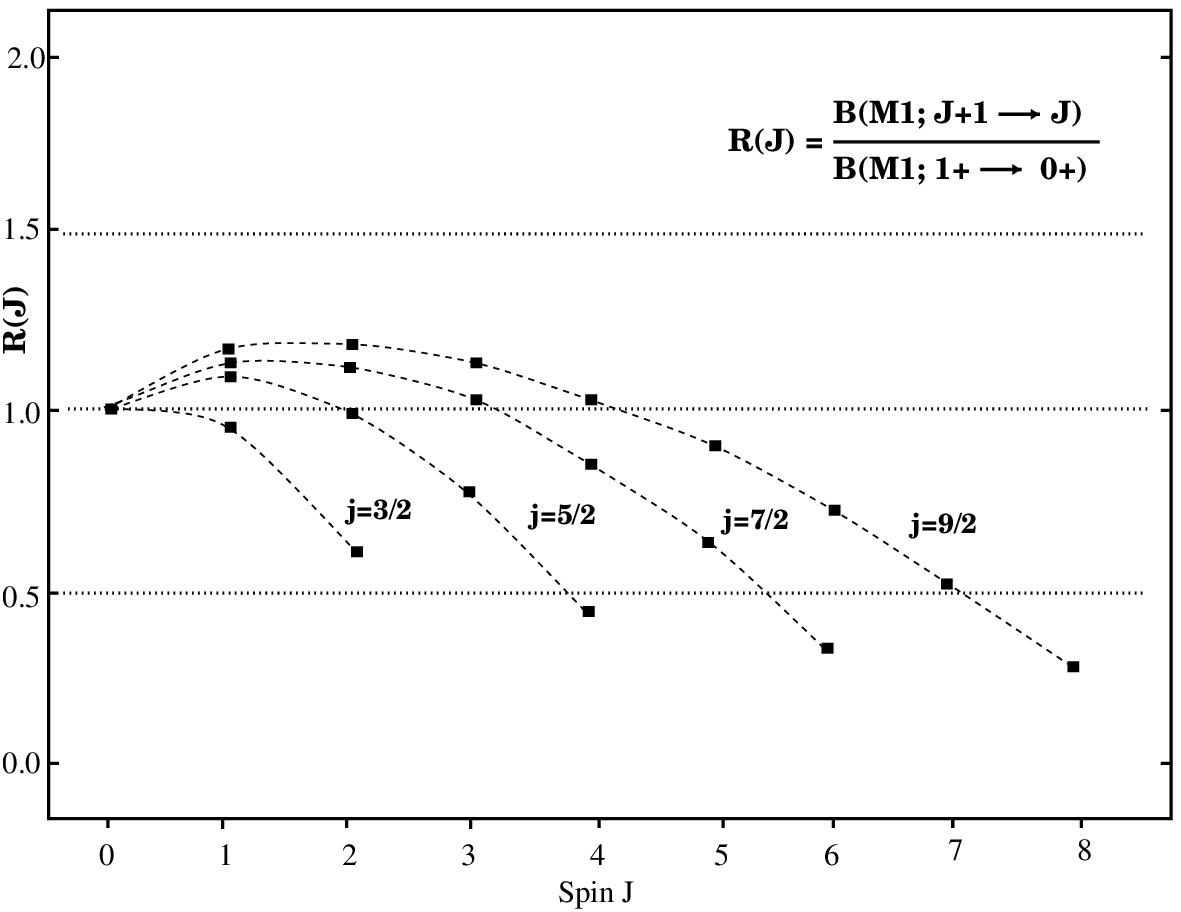,width=6in}}
\vspace*{10pt}
\caption{Ratio R of M1 transition strengths between QDC as a function of the 
total spin quantum number J plotted for different single particle 
orbitals $j=3/2,5/2,7/2,9/2$.}
\label{fig1}

\vspace{2cm}

\centerline{\epsfig{file=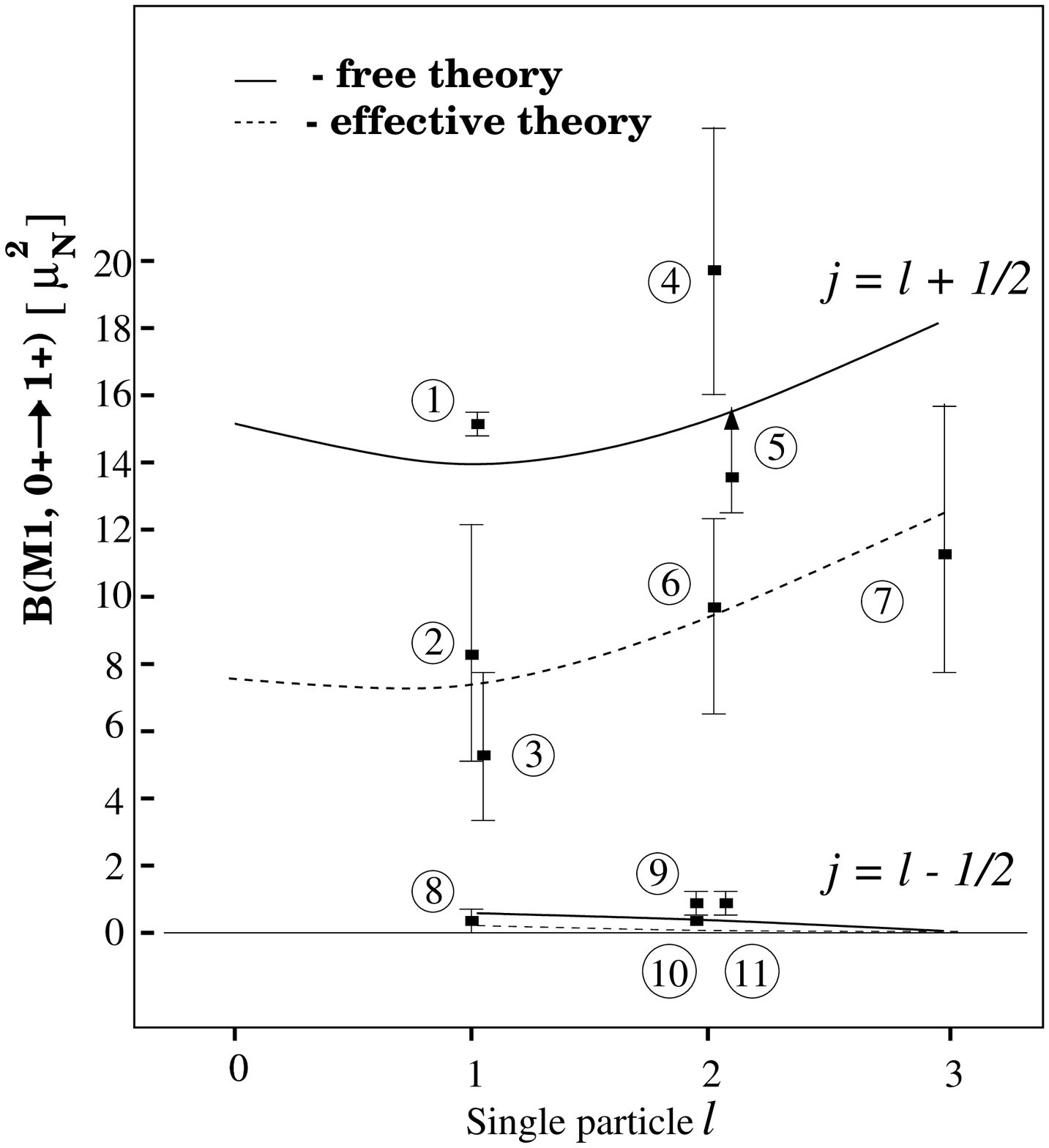,width=6in}}
\vspace*{10pt}
\caption{ The calculated  and experimental B(M1;$0^+ \rightarrow 1^+$) 
values are given as a function of a single particle angular 
momentum $l$. The results of calculations are shown for QDC j=l+1/2 branch 
(Eq.\ref{1bm1}) and for the j=l-1/2 branch (Eq.\ref{2bm1}).
The full lines correspond to the free theory ( free spin g-factors) and 
the dashed lines to the effective theory ( with quenching factor 
$\alpha_q$=0.7). The experimental data for different elements are labeled 
by numbers which are given in Table \ref{m1values}. The value for $^{22}$Na 
represents a lower limit. One expect the experimental values to lie between 
or in vicinity of the two lines  for both $j=l+1/2$ and $j=l-1/2$ cases.} 
\label{fig2}
\end{figure}

Other isovector transitions that involve states with the spin $J^\pi$ 
quantum number different from 
$J^\pi=0^+$ and $J^\pi=1^+$ ( $\Delta T=1$, $J+1 \rightarrow J$) in 
odd-odd N=Z 
nuclei  can be sizable in the strength with  the $1^+ \rightarrow 0^+$ 
transition strength. 
In the 2NSj one can derive the simple relation: 
\begin{eqnarray}
\label{mejj}
B(M1;J+1 \rightarrow J)={3(J+1)(2j+2+J)(2j-J) \over 4 j(j+1)(2J+3)}B(M1;1^+ \rightarrow 0^+).
\end{eqnarray}
This relation is valid for both $j=l+1/2$ and $j=l-1/2$ cases.  The 
dependence of the ratio $R=B(M1;J+1 \rightarrow J)/B(M1;1^+ \rightarrow 0^+)$ 
on the spin quantum number J of the final state for different single j is 
shown in Fig.\ref{fig1}. As it can be seen from Fig.\ref{fig1} the number of 
transitions sizable in strength with $1^+ \rightarrow 0^+$ transition 
increases with increasing j. Since the $B(M1,0^+ \rightarrow 1^+)$ value is 
large for the QDC, the $B(M1,J+1 \rightarrow J)$ values can be also large 
in this case. The QDC states with J=0,...,2j form the band of states 
connected by strong M1 transitions that is similar to the ``shears band'' in 
heavy nuclei \cite{Clark98}. The strong M1 transitions caused by the QDC in 
odd-odd N=Z nuclei and the strong M1 transitions related to the ``shears'' 
mechanism have similar noncollective nature.

But in the $j=l-1/2$ 
case one cannot expect large $B(M1;J+1 \to J)$ values 
because they are proportional to the small 
$B(M1;1^+ \to 0^+)$ value by a spin dependent 
proportionality factor, which is close to one.

The Eqs.(\ref{me1}) and (\ref{me2}) can be used also to derive 
within the single-$j$-shell approximation a unique 
formula for the B(M1) values for both $j=l+1/2$ and $j=l-1/2$ cases in 
terms of magnetic moments of neighboring odd-A nuclei.    
In the independent particle model, the magnetic dipole moment $\mu$ of a 
nucleon in an orbital ($nlj$) is given by the Schmidt values:
\begin{eqnarray}
\label{mj1+}
\mu_\rho(j=l+{1 \over 2})=\left[ g_l^\rho l+{g_s^\rho \over 2} \right]\mu_N,
\end{eqnarray}  
\begin{eqnarray}
\label{mj1-}
\mu_\rho(j=l-{1 \over 2})={j \over j+1} 
\left[ g_l^\rho (l+1)-{g_s^\rho \over 2} \right]\mu_N.
\end{eqnarray}
where $\rho$ = $\pi$ for proton and $\nu$ for neutron. 

The combination of the Eqs.(\ref{me1},\ref{me2}) with the  
Eqs.(\ref{mj1+},\ref{mj1-}) results in a simple relation between the M1  
strenghts of transitions between 2NSj states and magnetic moments of 
neighboring odd-mass nuclei:
\begin{eqnarray}
\label{bmu}
B(M1;(\pi j \times \nu j),0^+ \rightarrow (\pi j \times \nu j),1^+) = 
{3 \over 4\pi}{j+1 \over j}
\left[\rule{0mm}{5mm}\mu_\pi(j)-\mu_\nu(j)\right]^2. 
\end{eqnarray}
The spin-orbit interference is hidden now in $\mu_\pi(j)$ and  $\mu_\nu(j)$
quantities. 
Expression (\ref{bmu}) does not explicitly contain orbital and 
spin g-factors and therefore can help to explore the structure
of the yrast $0^+$ state and the $1^+$ state in odd-odd N=Z nucleus in 
an alternative way. 

In contrast to the isovector M1 transitions discussed above, 
the magnetic dipole moments 
in odd-odd nuclei are generated by the isoscalar (IS) part 
\begin{equation}
{\bf T}_{IS}(M1)=\left[{g_p^l+g_n^l \over 2}({\bf L}_p+{\bf L}_n)
+{g_p^s+g_n^s \over 2}({\bf S}_p+{\bf S}_n)\right]{\mu_N \over \hbar} 
\end{equation}  
of the M1 operator. 
We consider now magnetic dipole moments of odd-odd nuclei 
in the 2NSj approximation. The expressions for the M1 moment in a state with 
angular momentum quantum number J can be written in the following way:
\begin{eqnarray}
\label{mj+}
\mu(j=l+{1 \over 2})={J \over 2l+1}
\left[(g_p^l+g_n^l)l+{g_p^s+g_n^s \over 2} \right]\mu_N,
\end{eqnarray}  
and 
\begin{eqnarray}
\label{mj-}
\mu_\rho(j=l-{1 \over 2})={J \over 2l+1} 
\left[(g_p^l+g_n^l)(l+1)-{g_p^s+g_n^s \over 2}\right]\mu_N. 
\end{eqnarray}
The partial cancellation of spin and orbital parts in the $j=l-1/2$ case 
 is also obvious from Eq.(\ref{mj-}). However, the $\mu$ values are not very sensitive to the spin 
part due to the small value of the sum of proton and neutron spin g-factors.

\section{Discussion}

In this section we will confront the simple formulae 
from above with the data.
Using Eqs.(\ref{me1},\ref{me2}) and  free g-factors 
($g_p^l=1.0$, $g_n^l=0.0$, $g_p^s=5.5857$ and $g_n^s=-3.8263$) we obtain for 
the 
$B(M1;(\pi j \otimes \nu j); J^\pi=0^+ \rightarrow (\pi j \otimes \nu j) J^\pi=1^+)$ values the expressions:  

\begin{eqnarray}
\label{1bm1}
B(M1; 0^+ \rightarrow 1^+) = {3\over 4\pi}{j+1 \over j} 
\left[ \rule{0mm}{5mm} l + 4.706 \right]^2\mu_N^2, \hspace{0.3cm} \mbox{for}\hspace{0.3cm}  
j=l+{1 \over 2} 
\end{eqnarray}
and 
\begin{eqnarray}
\label{2bm1}
B(M1;0^+ \rightarrow 1^+) = {3\over 4\pi}{j \over j+1} 
\left[\rule{0mm}{5mm}l - 3.706 \right]^2\mu_N^2, \hspace{0.3cm} \mbox{for}\hspace{0.3cm} 
j=l-{1 \over 2}. 
 \end{eqnarray}

The B(M1) values from Eqs.(\ref{1bm1},\ref{2bm1}))are plotted with solid 
curves in Fig. \ref{fig2} as functions of the single particle orbital 
angular momentum $l$. The deuteron belongs to the $j=l + 1/2$ 
branch. Only the spin part would contribute to the $B(M1)$ value 
if the $J^\pi=0^+_{T=1}$ state of the deuteron were bound. The calculated 
B(M1) value for a modeled bound state is very large: 
$B(M1; (\pi s_{1/2}\nu s_{1/2}),J=0^+ \rightarrow  (\pi s_{1/2}\nu s_{1/2}),
J=1^+)$=15.86 $\mu_N^2$. 
Other $B(M1)$ values calculated from Eqs.(\ref{1bm1},\ref{2bm1}) assuming 
reasonable single particle orbitals are given in Table \ref{m1values} 
together with the corresponding experimental $B(M1)$ values.

For $j=l+1/2$ cases  Eq.(\ref{1bm1}) yields large B(M1) values sizable with 
the 
B(M1;$0^+ \rightarrow 1^+)$ for the deuteron.  Therefore the large  
B(M1;$0^+ \rightarrow 1^+$) 
values in odd-odd N=Z nuclei can be considered an indication of the QDC.     
The strong $M1$ transitions in $^6$Li and $^{18}$F are the best 
examples for transitions between QDC. 
In $^{22}$Na a large lower limit for the total $B(M1;0^+\to 1^+)$ 
value is obtained from summing up the $M1$ strengths from 
three transition fragments (see Table \ref{m1values}). This value agrees with 
the estimate for a $B(M1)$ value between QDC, too.
 
For the four odd-odd $N=Z$ nuclei $^{10}$B, $^{14}$N, $^{26}$Al 
and $^{42}$Sc large $B(M1)$ values are known, as well. 
These values are, however, by a factor of about two smaller than 
the corresponding QDC estimates with free $g$-factors. 
Deviations of the data from the simple expressions (\ref{1bm1},\ref{2bm1}) 
can be attributed to configuration mixing \cite{Arima}, which are neglected in 
the simple 2NSj.

Configuration mixing can be taken into account to a certain extent 
by using quenched spin $g$-factors $g_s = \alpha_q g_s^{\rm free}$ 
with a quenching factor $\alpha_q$. 
We have computed effective $B(M1)$ values with $\alpha_q = 0.7$  for both 
cases $j=l+1/2$ and $j=l-1/2$ from Eqs.((\ref{1bm1},\ref{2bm1})). 

The results are included in Table \ref{m1values} and plotted with dashed 
curves in Fig.\ref{fig2}.
The effective $B(M1)$ values agree with the data from $^{10}$B, 
$^{14}$N, $^{26}$Al and $^{42}$Sc. 
These agreements with the estimates using quenched $g$-factors 
indicate that a precise quantitative understanding 
requires larger scale shell model calculations. The main mechanism is, 
however, understood already in the simple 2NSj.

Small experimental isovector $B(M1)$ values were found in $^{14}$N, $^{30}$P, 
$^{34}$Cl and $^{38}$K. Small $B(M1)$ values are calculated in the 2NSj for 
the $j=l-1/2$ case, regardless whether free $g$-factors or 
quenched $g$-factors are used.
The best examples for an isovector $M1$ transition between the 
$0^+_1$ state and the $1^+_1$ state, which almost vanishes 
due to the cancellation of the orbital and the spin parts, 
are observed in $^{34}$Cl and $^{38}$K. 
This cancellation can reduce the $B(M1)$ value by a factor of more 
than 20 in comparison to the large {\em quasi-deuteron} $M1$ 
transitions.
Such a drastic difference between {\em quasi-deuteron} transitions and  
{\em non-quasi-deuteron} transitions can be qualitatively well 
understood within  the 2NSj approximation.

Particularly interesting cases are $^{14}$N and  $^{30}$P.
In $^{14}$N two kinds of transitions coexist at low energy.
The transition from the $0^+_1$ state to the $1^+_1$ ground state 
is very weak  and can be related to the $j=l-1/2$ ($p_{1/2}$ shell ) case. 
The next $1^+_2$ state  at 3.948 MeV is also low-lying and connected with the 
$0^+_1$ state by a strong M1 transition. This transition can be considered as 
a {\em quasi-deuteron} transition  ($j=l+1/2$ case) in the $p_{3/2}$ shell.
It means that both $\pi(1p_{1/2}^1)\times\nu(1p_{1/2}^1)$ and 
$\pi(1p_{3/2}^{-1})\times\nu(1p_{3/2}^{-1})$ components 
must be present in the $0^+_1$ wave function with amplitudes 
smaller than one. 
This fact may explain  why 
the experimental $B(M1;0^+_1 \rightarrow 1^+_2)$ transition 
strength is smaller than the estimated one.  
Similarly one can explain the fact 
that the experimental B(M1) value for the $0^+_1 \rightarrow 1^+_1$ transition 
in $^{30}$P is larger than the calculated value for the supposed 
$\pi(1d_{3/2}^1)\times\nu(1d_{3/2}^1)$ $j=l-1/2$ configuration 
of the $0^+_1$ and 
$1^+_1$ states. A small fragment of the {\em quasi-deuteron}
$(\pi(2s_{1/2}^1)\times\nu(2s_{1/2}^1);0^+_1) \rightarrow 
(\pi(2s_{1/2}^1)\times\nu(2s_{1/2}^1);1^+_1)$ transition enhances the 
$0^+_1 \rightarrow 1^+_1$ transition.

Let us now discuss the relation (\ref{bmu}) between isovector 
$M1$ transitions in the 2NSj and the Schmidt values for magnetic 
dipole moments, which is an exact equation in the simple 
single-$j$-shell approximation. 
As it was discussed above $B(M1)$ values in odd-odd $N=Z$ nuclei 
can differ from the pure 2NSj estimates due to configuration mixing. 
This could be partly taken into account by using quenched 
spin $g$-factors. 
On the other hand, configuration mixing can lead also to a deviation 
of $M1$ moments in odd-$A$ nuclei from the Schmidt values, which were 
used to eliminate the $g$-factors in Eq. (\ref{bmu}). 
It is, therefore, very interesting to investigate to what 
extent Eq. (\ref{bmu}) can be used to predict isovector $B(M1)$ 
values in odd-odd $N=Z$ nuclei from magnetic dipole moments 
in neighboring odd-$A$ nuclei. 
For this purpose we replace $\mu_\pi(j)$ and $\mu_\nu(j)$ in 
Eq.(\ref{bmu}) with the corresponding
experimental values. 
The experimental $\mu_\pi$ and $\mu_\nu$ values are the 
magnetic moments of the ground states $J^\pi=j^\pi$ in the neighboring 
odd-proton and odd-neutron nuclei, respectively. 
Comparing the 
B(M1;$0^+_1 \rightarrow 1^+_1$) value calculated in this way with the 
corresponding experimental value we can conclude about the structure of the 
$0^+$ and $1^+$ states. 

As an example, let us consider the nucleus $^{42}$Sc. 
This nucleus has one 
neutron and one proton occupying the $1f_{7/2}$ shell above the even-even 
$^{40}$Ca core. 
The experimental magnetic dipole moments of the $J^\pi=7/2^-$ ground 
states in the neighboring nuclei $^{41}$Sc and $^{41}$Ca 
are 5.535 $\mu_N$ and -1.595 $\mu_N$, respectively. Substituting 
these values in Eq.(\ref{bmu}) we get for $^{42}$Sc 
B(M1;$0^+ \rightarrow 1^+$)=15.62 $\mu_N^2$. 
Comparing this value with the experimental value of 10(4) $\mu_N^2$, we can 
conclude that the wave functions of the $0^+_1$ ground state and 
the excited $1^+_1$ state in $^{42}$Sc are dominated by the  
$(7/2^-_\pi)\times(7/2^-_\nu)$ component, where the 
$J^\pi_\rho=7/2^-_\pi$ and the $J^\pi_\rho=7/2^-_\nu$ states 
are the ground states of $^{41}$Sc and $^{41}$Ca, respectively.

The B(M1) values estimated from $M1$ moments in neighboring 
odd-$A$ nuclei and the 
corresponding experimental data are shown in Table \ref{comp} for other 
odd-odd N=Z nuclei. 
Here, we consider for all nuclei only the lowest $0^+ \to 1^+$ 
transitions. 
The estimated B(M1) value for the $^{14}$N nucleus 
is larger than the experimental one. This supports our schematic 
explanation of the 
mixing of the QDC states with the states formed by a 
proton-neutron pair in the 
$j=l-1/2$ orbital. The results for the nucleus $^{30}$P are also interesting. 
They can be interpreted in the following way: The $0^+_1$ state and the 
$1^+_1$ state  contain a large $(1/2^+_\pi)\times(1/2^+_\nu)$ component. But 
the $1/2^+_\pi$ ground state of the nucleus  $^{29}$P  and the $1/2^+_\nu$ 
ground state of $^{29}$Si cannot be pure $2s_{1/2}$ states because 
the corresponding magnetic moments differ very much from the 
Schmidt values. 
These states should have more complicated 
structures that involve at least the $d_{3/2}$ and $d_{5/2}$ orbitals. 
It explains partially why there are no low-lying pure QDC states 
in $^{30}$P.
The estimated B(M1) values for other nuclei are in 
rough agreement with the data. 
We conclude that the
structures of the $0^+_1$ and the $1^+_1$ states can often be well 
approximated by the direct product $(J^\pi_p)\times(J^\pi_n)$ of 
the ground states of 
the corresponding (see Table \ref{comp}) odd-Z and odd-N nuclei.

We do not discuss here the nuclear magnetic dipole moments. 
This has been done recently \cite{Ron98}. 
We would like only to note that the isoscalar $M1$ moments are 
less sensitive to the spin part of the M1 m.e. 
$[$see Eqs.(\ref{mj+},\ref{mj-})$]$ than the isovector $M1$ transitions. 
This well known fact is due to the relatively small 
value of the isoscalar spin $g$-factor $(g_p^s+g_n^s)/2$ 
(0.88 for free spin g-factors) in comparison 
to the isovector value $(g_p^s-g_n^s)/2$ (4.706 for free spin g-factors). 
Therefore, the constructive or destructive interference between 
the spin part and the orbital part for the cases with $j=l+1/2$ 
and with $j=l-1/2$ are less pronounced for the isoscalar $M1$ moments 
than for the isovector $B(M1)$ values. 
The $\mu$ values for the $j=l+1/2$ and the $j=l-1/2$ branches differ by 
approximately a factor of 2 for $j=1/2$ and become almost equal at 
large j values. 
This is in an agreement with experiment. 
It is more difficult to see the difference between the two branches 
studying the magnetic dipole moments in odd-odd N=Z nuclei. 
This difference is easier to observe by 
analyzing the B(M1) values regardless of their large 
experimental errors. 
Therefore, isovector B(M1) values can be a more sensitive tool 
for the investigation of quasi-deuteron configurations. 

\section{Conclusions}

Isovector magnetic dipole transitions between low-lying $0^+$ and 
$1^+$ states in odd-odd N=Z nuclei were studied within 
the simple core+two-nucleon single-j-shell model. 
Analytical expressions for isovector B(M1) values in odd-odd N=Z nuclei 
were derived. 
Low-lying states in odd-odd $N=Z$ nuclei with a proton-neutron pair in a 
$j=l+1/2$ shell were considered as quasi-deuteron configurations. 
These cases are characterized by large B(M1) transition strenghts 
caused by coherent contributions of the orbital and spin parts to 
the total strength. 
The large $B(M1)$ values were interpreted as direct indications of 
QDC in the states which are connected by these strong transitions. 
Incoherent contribution of the spin 
and orbital parts to the total transition strength for the states formed by 
the proton-neutron pair in the $j=l-1/2$ shell strongly reduces the B(M1) 
values. 
The B(M1) values for the 
low-lying states in odd-odd nuclei can be predicted knowing only 
the single particle $j$ quantum number for the orbital occupied with the 
proton-neutron pair. 
Low-lying QDC states can be expected in the 
$1f_{7/2}$ nuclei $^{46}$V, $^{50}$Mn and $^{54}$Co. Weaker transitions 
can be expected in $^{58}$Cu nucleus ( $2p_{3/2}$ shell). In the odd-odd N=Z 
nuclei where the $1f_{5/2}$ and $2p_{1/2}$ valence orbitals configurations 
become dominant for low-lying states one can not expect to observe strong M1 
transitions between these low-lying states.  Only in the nuclei where the 
$1g_{9/2}$ shell plays  important role (for example $^{82}$Nb nucleus) one 
can expect again the low-lying QDC states connected by strong M1 transitions 
(see Table \ref{comp}).      
It is very interesting to identify the QDC in the heavier odd-odd N=Z 
nuclei and to check how well they fit into the picture. Recent experiments 
on low-lying states in the odd-odd N=Z nuclei $^{46}$V \cite{Fries98} and 
$^{54}$Co \cite{Schneider99} already give some preliminary indications   
\footnote{The measured branching ratios and multipole mixing ratios for some 
transitions and isospin symmetry with the neighboring nuclei require strong 
M1 transitions.} about the  existence of QDC in these nuclei with the 
$f_{7/2}$ shell being the valence shell.
We have related the B(M1) values in odd-odd N=Z nuclei with the 
magnetic moments of neighboring odd-A nuclei. 
The established simple 
connection can provide additional information on the structure of the 
$0^+_1,T=1$ state and the $1^+_1,T=0$ state. 

\section{Acknowledgment}

We gratefully acknowledge valuable discussions with C.~Frie\ss{}ner, 
A.~Schmidt, I.~Schneider, Dr.~J.~Eberth, Prof.~T.~Otsuka, Prof.~A.~Gelberg, 
Dr.~R.~S.~Chakrawarthy and Dr.~L.~E\ss{}er. 
One of us (R.V.J.) thanks the Universit\"at zu K\"oln 
for a Georg Simon Ohm guest professorship.

\begin{table}
\caption{Experimental [27-29] and calculated B(M1) values for odd-odd N=Z nuclei.
 M1 transitions with less than 5$\%$ of the total strength are omitted.
 In the column ``free theory'' results of calculations  for the free spin 
g-factors are shown  and in the column ``eff.theory'' -- for the 
effective spin g-factors $g_s^{\rm eff}$=0.7$g_s^{\rm free}$.}
\label{m1values}
\begin{center}
\begin{tabular}{c|c|c|c|c|c|ccc}
 N & Nucleus & used & $E_x^{exp}(0^+)$, & $E_x^{exp}(1^+)$,&
B(M1;$0^+\rightarrow 1^+$),& \multicolumn{3}{c}{ $\sum$ B(M1;$0^+\rightarrow 1^+$),$[\mu_N^2]$} \\  
\cline{7-9}
 &  & configuration & [MeV] & [MeV] & $[\mu_N^2]$ & exp. & free theory &  eff. theory  \\ 
\hline
  &     deuteron          & $\pi1s_{1/2}\nu1s_{1/2}$ & - & 0 &  & & 15.86 & 7.77  \\
1 & $^6_3$Li$_3$          &  $\pi1p_{3/2}\nu1p_{3/2}$ & 3.562 & 0 & 15.4(4)& 15.4(4) & 12.96 & 7.34  \\
2 & $^{10}_5$B$_5$        &  $\pi1p_{3/2}^{-1}\nu1p_{3/2}^{-1}$ & 1.740 & 0.718& 7.5(32) & 8.1(33) & 12.96 & 7.34  \\
  &  & & 1.740 & 2.154 & 0.59(5) & &  &  \\
3 & $^{14}_7$N$_7$        &  $\pi1p_{3/2}^{-1}\nu1p_{3/2}^{-1}$ &  2.312 & 3.948 & 5($^{+3}_{-2}$) & 5($^{+3}_{-2}$) & 12.96 & 7.34 \\
4 & $^{18}_9$F$_9$        &  $\pi1d_{5/2}\nu1d_{5/2}$ & 1.041 & 0 & 20(4)& 20(4) & 15.18 & 9.37 \\
5 & $^{22}_{11}$Na$_{11}$ &  $\pi1d_{5/2}^3\nu1d_{5/2}^3$ & 0.657 & 0.583 & 5.0(3)&$>$12.5 & 15.18 & 9.37  \\
 & & & 0.657 & 1.936 & 4.4(10) && & \\
 & & & 0.657 & 3.943 & $>$4.4  && & \\
6 & $^{26}_{13}$Al$_{13}$ &  $\pi1d_{5/2}^{-1}\nu1d_{5/2}^{-1}$ & 0.228 & 1.057 & 8(2) & 9.4(30) & 15.18 & 9.37  \\
 & & & 0.228 & 1.850 & 0.8(8) & & & \\
 & & & 0.228 & 3.723 & 0.6(2) & & & \\
7& $^{42}_{21}$Sc$_{21}$ &  $\pi1f_{7/2}\nu1f_{7/2}$ & 0 & 0.611 & 11(4)& 11(4) & 18.23 & 12.16  \\ \hline
8 & $^{14}_7$N$_7$        &  $\pi1p_{1/2}\nu1p_{1/2}$ & 2.312 & 0 & 0.05(2)& 0.05(2)&  0.58 &  0.13   \\
9 & $^{30}_{15}$P$_{15}$  &  $\pi1d_{3/2}\nu1d_{3/2}$ & 0.677 & 0 & 1.3(1) & 1.3(1) & 0.42 &  0.012  \\
10 & $^{34}_{17}$Cl$_{17}$ &  $\pi1d_{3/2}\nu1d_{3/2}$ & 0  & 0.461 & 0.23(2) & 0.23(2) & 0.42 &  0.012 \\
11 &$^{38}_{19}$K$_{19}$  &  $\pi1d_{3/2}^{-1}\nu1d_{3/2}^{-1}$ & 0.130 & 0.460 &0.47(4) &  1.17(24) & 0.42 &  0.012   \\
 & & & 0.130 & 1.698 & 0.7(2) & & & \\
\end{tabular}
\end{center}
\end{table}

\begin{table}
\caption{Experimental [30,31] magnetic dipole moments of  odd-N and odd-Z nuclei, 
experimental [27-29] and estimated (see Eq.\ref{bmu}) 
B(M1;$0^+_1 \rightarrow 1^+_1$) values for odd-odd nuclei.}
\label{comp}
\begin{center}
\begin{tabular}{ccc|ccc|c|cc}
\multicolumn{3}{c|}{odd-Z} & \multicolumn{3}{c|}{odd-N} & Odd-odd & 
\multicolumn{2}{c}{B(M1;$0^+_1 \rightarrow 1^+_1$),$[\mu_N^2]$}  \\ 
\cline{1-6}\cline{8-9}
Nucleus & $J^\pi_p$ & $\mu_p^{exp},[\mu_N]$ & Nucleus & $J^\pi_n$ & $\mu_n^{exp},[\mu_N]$ & nucleus & exp. & Eq.(\ref{bmu}) \\ 
\hline
$^{11}_5$B$_6$ &$3/2^-$ & 2.689 & $^{11}_6$C$_5$ &$3/2^-$ & -0.964(1) & 
$^{10}_5$B$_5$ & 7.5(32) & 5.32 \\
$^{13}_7$N$_6$ & $1/2^-$& -0.322 & $^{13}_6$C$_7$ &$1/2^-$ & 0.702 & 
$^{14}_7$N$_7$& 0.05(2) & 0.75 \\
$^{17}_9$F$_8$ &$5/2^+$& 4.722 & $^{17}_8$O$_9$ &$5/2^+$ & -1.894 & 
$^{18}_9$F$_9$ & 20(4) & 14.65  \\
$^{21}_{11}$Na$_{10}$ & $3/2^+$ & 2.836 & $^{21}_{10}$Ne$_{11}$ &$3/2^+$ 
&  -0.662 &$^{22}_{11}$Na$_{11}$& 5.0(3) & 4.87  \\
$^{25}_{13}$Al$_{12}$ & $5/2^+$& 3.646 & $^{25}_{12}$Mg$_{13}$ & $5/2^+$ & -0.855 &$^{26}_{13}$Al$_{13}$& 8(2) & 6.78 \\
$^{29}_{15}$P$_{14}$& $1/2^+$& 1.235 & $^{29}_{14}$Si$_{15}$ &$1/2^+$ & -0.555 &$^{30}_{15}$P$_{15}$ & 1.3(1) & 2.33 \\
$^{33}_{17}$Cl$_{16}$ &$3/2^+$ & 0.752 & $^{33}_{16}$S$_{17}$ &$3/2^+$ & 0.644 & $^{34}_{17}$Cl$_{17}$ & 0.23(2) & 0.005 \\
$^{37}_{19}$K$_{18}$ & $3/2^+$ & 0.203  & $^{37}_{18}$Ar$_{19}$ & $3/2^+$ & 1.145 &$^{38}_{19}$K$_{19}$ & 0.47(4) & 0.35  \\
$^{41}_{21}$Sc$_{20}$ & $7/2^-$ & 5.535  & $^{41}_{20}$Ca$_{21}$ & $7/2^-$& -1.595 &$^{42}_{21}$Sc$_{21}$ & 11(4) & 15.62  \\
$^{49}_{23}$V$_{26}$ &$7/2^-$ & 4.47(5) & $^{45}_{22}$Ti$_{23}$& $7/2^-$& -0.095(2) &$^{46}_{23}$V$_{23}$ & - & 6.40  \\
$^{51}_{25}$Mn$_{26}$ & $5/2^-$& 3.5683 & $^{47}_{22}$Ti$_{25}$ & $5/2^-$ & -0.788 & $^{50}_{25}$Mn$_{25}$ &  - & 5.82 \\
$^{55}_{27}$Co$_{28}$ & $7/2^-$& 4.822(3) & $^{47}_{20}$Ca$_{27}$ & $7/2^-$ & -1.380(24) & $^{54}_{27}$Co$_{27}$ &  - & 11.82 \\
$^{61}_{29}$Cu$_{32}$ & $3/2^-$& 2.14(4) & $^{57}_{28}$Ni$_{29}$ & $3/2^-$ & -0.798(1) &$^{58}_{29}$Cu$_{29}$ &-& 3.44 \\
$^{89}_{41}$Nb$_{48}$ & $9/2^+$ & 6.216(5) & $^{89}_{40}$Zr$_{49}$ & $9/2^+$ & -1.076(20) & $^{82}_{41}$Nb$_{41}$ & - & 15.52 \\
\end{tabular}
\end{center}
\end{table}

\end{document}